\documentclass[pdflatex,sn-mathphys-num,
]{sn-jnl}

\usepackage{graphicx}
\usepackage{amsmath}
\usepackage{amssymb}
\usepackage{xcolor}
\usepackage{bbm}

\makeatletter
\let\saved@includegraphics\includegraphics
\AtBeginDocument{\let\includegraphics\saved@includegraphics}
\renewenvironment*{figure}{\@float{figure}}{\end@float}
\makeatother

\newcommand{\mmu}{\text{\textmu}}
\newcommand{\ee}{\textrm{e}}
\newcommand{\ii}{\textrm{i}}

\begin{document}

\title{Zeptojoule calorimetry}

\author*[1]{\fnm{András Márton} \sur{Gunyhó}}\email{andras.gunyho@aalto.fi}
\author[1]{\fnm{Kassius} \sur{Kohvakka}}
\author[1]{\fnm{Qi-Ming} \sur{Chen}}
\author[1]{\fnm{Jean-Philippe} \sur{Girard}}
\author[1]{\fnm{Roope} \sur{Kokkoniemi}}
\author[1,2]{\fnm{Wei} \sur{Liu}}
\author*[1,3]{\fnm{Mikko} \sur{Möttönen}}\email{mikko.mottonen@aalto.fi}

\affil[1]{\orgdiv{QCD Labs, QTF Centre of Excellence, Department of Applied Physics}, \orgname{Aalto University}, \orgaddress{\street{P.O.~Box 13500}, \city{Espoo}, \postcode{FIN-00076},
\country{Finland}%
}}

\affil[2]{\orgname{IQM Quantum Computers}, \orgaddress{\street{Keilaranta 19},
\city{Espoo},
\postcode{02150}
\country{Finland}%
}}

\affil[3]{\orgdiv{QTF Centre of Excellence}, \orgname{VTT Technical Research Centre of Finland Ltd.}, \orgaddress{\street{P.O.~Box 1000},
\postcode{02044 VTT},
\country{Finland}%
}}

\abstract{\vspace*{-5pt}
    The measurement of energy is a fundamental tool used, for example, in exploring the early universe\cite{planck_2018_results,Wiedner2021}, characterizing particle decay processes\cite{ranitzsch2020metrommc,pagnanini2023array}, as well as in quantum technology and computing\cite{crain2019highspeed,clerk2020hybrid,casariego2023propagating,chou2024dualrail}.
    Some of the most sensitive energy detectors are thermal, i.e., bolometers and calorimeters\cite{Mccammon2005}, which operate by absorbing incoming energy, converting it into heat, and reading out the resulting temperature change electrically using a thermometer.
    Extremely sensitive calorimeters, including transition edge sensors\cite{ullomReviewSuperconductingTransitionedge2015,delucia2024transition}, magnetic microcalorimeters\cite{kempfPhysicsApplicationsMetallic2018,kim2023cryogenic} and devices based on 2D conductors such as graphene\cite{samaha2024graphene,dibattista2024infraredgraphene,huang2024graphenecalorimetric}, have been shown to reach impressive energy resolutions of 17.6~zJ\cite{karasik2012infrared}.
    Very recently superconductor--normal-conductor--superconductor~(SNS) radiation sensors with metallic\cite{Roope_JPA} and graphene\cite{Kokkoniemi2020} absorbers have resulted in predictions of full-width-at-half-maximum (FWHM) energy resolutions of 0.75~zJ and 0.05~zJ = 71~GHz$\times h$, respectively, where $h$ is the Planck constant.
    However, since these estimates are only mathematically extracted from steady-state noise and responsivity measurements, no calorimetry reaching single-zeptojoule energy resolution or beyond has been demonstrated.
    Here, we use a metallic SNS sensor to measure the energy of 1-µs-long 8.4-GHz microwave pulses with a FWHM energy resolution finer than (0.95~$\pm$~0.02)~zJ = (5.9~$\pm$~0.12)~meV, corresponding to 170 photons at 8.4~GHz.
    The techniques of this work, combined with the graphene-based sensors of Ref.~\cite{Kokkoniemi2020}, provide a promising path to real-time calorimetric detection of single photons in the 10 GHz range.
    Such a device has potential in operating as an accurate measurement device of quantum states such as those of superconducting qubits\cite{Gunyho2024}, or used in fundamental physics explorations including quantum thermodynamics\cite{Pekola2015,myers2022quantumthermodynamic}, and the search for axions\cite{CAPPARELLI201637,pankratov2022microwave}.
}

\newgeometry{top=2cm}

\maketitle

\restoregeometry

\section*{Introduction}

The detection of weak electromagnetic signals is of great interest in a broad range of scientific and practical applications.
Indeed, leaps in the scientific understanding of nature have often been driven by new measurement data enabled by improved sensors.
This has spurred the development of several kinds of ultrasensitive radiation sensors operating at cryogenic temperatures.
Currently, it is possible to detect individual microwave photons below $10\,\mathrm{GHz}$ using, for example, superconducting qubits\cite{Qubits_BW_10MHz_biblio,balembois2024cyclically}, current-biased Josephson junctions\cite{Chen_PRL_detector_JJ,golubev2021JJ,pankratov2024JJthermal} or nonlinear oscillators\cite{petrovnin2024photondetection}.
However, such detectors are usually not energy resolving, and they may suffer from narrow bandwidths for the signal to be detected, typically on the order of a few megahertz.

Detecting single photons in an energy-resolving manner over a broad frequency band is possible using sensors where the incoming photons break Cooper pairs in a superconductor, which is the basis for the operating principle of kinetic inductance detectors (KIDs)\cite{KIDs_microwave_biblio,ulbricht2021applicationsMKID}, superconducting nanowire single-photon detectors\cite{you2020snspd,esmaeilzadeh2021snspd}, and quantum capacitance detectors (QCDs)\cite{echternach2018single,echternach2021array}.
Recently, single $25\,\mathrm{\mmu m}$ ($7.95\,\mathrm{zJ}$) photons have been resolved using a KID\cite{day2024singlephotonkid}, while single-photon detection down to energies of $1.5\,\mathrm{THz}\times h \approx 0.99\,\mathrm{zJ}$ has been demonstrated using QCDs\cite{echternach2018single,echternach2021array}.

The fundamental limit of the energy measured by pair-breaking detectors is set by the energy gap of the absorber material, which is roughly $100\,\mathrm{GHz} \times h$ for Al for example.
On the other hand, thermal detectors are only limited by thermal fluctuations owing to their coupling to a heat bath\cite{karimi_reaching_2020}, which has been predicted to enable energy resolutions as fine as $2\,\mathrm{GHz}\times h$\cite{paolucci2020JES}.
The most sensitive calorimeters reported to date are graphene-based devices able to detect single $1550\,\mathrm{nm}$ ($193 \,\mathrm{THz} \times h = 128\,\mathrm{zJ}$) photons, and a Ti transition edge sensor, which has been shown to have a resolution of $26.6 \,\mathrm{THz} \times h = 17.6\,\mathrm{zJ}$ for $8\,\mathrm{\mmu m}$ photons\cite{karasik2012infrared}.
Despite these advancements, calorimetry in the microwave regime ($< 300\,\mathrm{GHz}$), relevant for example for the axion search, remains a major outstanding challenge.

Recently, superconductor--normal-conductor--superconductor~(SNS) radiation sensors with metallic\cite{Roope_JPA} and graphene\cite{Kokkoniemi2020,lee2020graphene} absorbers have demonstrated record-low noise equivalent powers (NEPs) on the order of tens of $\mathrm{zW / \sqrt{Hz}}$, when they were operated as continuous power meters, i.e.,~bolometers.
Based on these NEP values, it was predicted that these sensors should have full-width half-maximum (FWHM) energy resolutions of $0.75\,\mathrm{zJ}$ and $0.05\,\mathrm{zJ} = 71 \mathrm{GHz} \times h $ for the metallic and graphene absorber, respectively.
However, calorimetric photon detection below one zeptojoule has not been demonstrated to date.

In this work, we utilize a metallic SNS sensor\cite{Govenius2014, Roope_JPA} as a zeptojoule calorimeter.
First, we characterize the noise equivalent power of the sensor in the bolometric mode, and subsequently record individual traces of the detector signal in the time domain while applying short microwave pulses in the zeptojoule regime at the input of the sensor.
By employing a matched filter on the traces, we find a FWHM energy resolution finer than $0.95\,\mathrm{zJ} \approx 1.4\,\mathrm{THz} \times h$.
This is on par with state of the art energy resolution for cryogenic detectors achieved by QCDs, while outperforming, to the best of our knowledge, the best calorimeters reported in the literature by an order of magnitude.
These results enable photon counting in the terahertz regime and provide a solid stepping stone to the calorimetric detection of microwave photons.

\section*{Device operation}

\begin{figure}
    \centering
    \includegraphics[
    width=3.5in
    ]{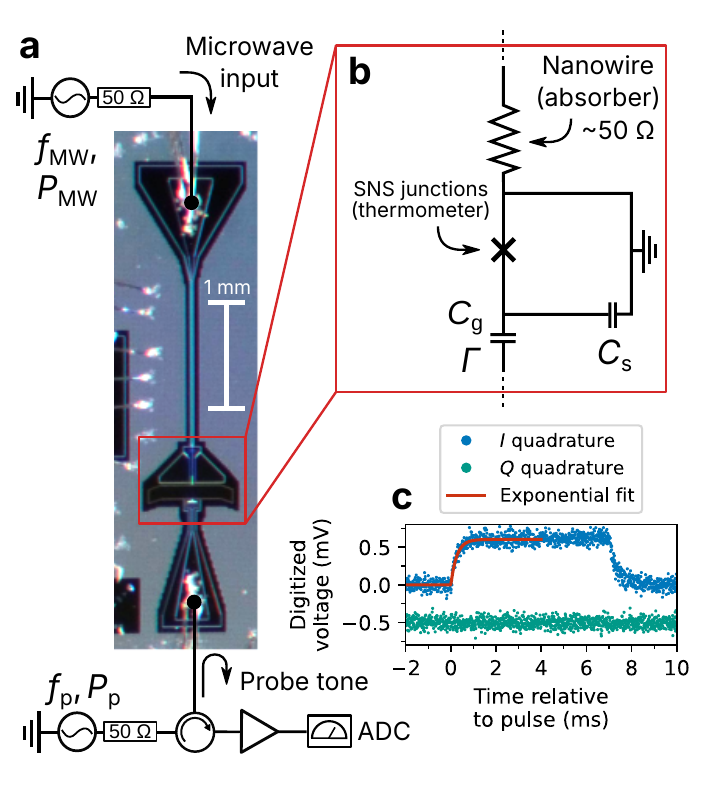}
    \caption{\textbf{Su\-per\-con\-duc\-tor--normal-conductor--su\-per\-con\-duc\-tor~(SNS) radiation sensor and its operating principle.}
    \textbf{a}, Optical microscope image of the SNS sensor chip together with a simplified measurement setup.
    \textbf{b}, Equivalent circuit diagram of the absorber and probe circuit of the sensor.
    \textbf{c}, Example of response of the sensor upon absorption of a $7$-$\mathrm{ms}$-long $8.40\,\mathrm{GHz}$ microwave pulse.
    The two quadratures of the digitized probe signal are shown (blue and green dots), as well as an exponential fit to the rising edge (red curve), used to extract the thermal time constant.
    The quadrature-phase trace is offset by $-0.5\,\mathrm{mV}$ for clarity.
}
    \label{fig:operating_principle}
\end{figure}

    Our SNS radiation sensor, similar to the device discussed in Ref.~\cite{Roope_JPA}, consists of a microwave absorber which is thermally coupled to a thermometer, as shown in Figs.~\ref{fig:operating_principle}a,b.
    The absorber is a micron-long resistive AuPd normal-metal nanowire, which is essentially a resistor and impedance-matched with the input microwave transmission line.
    The thermometer consists of several SNS Josephson junctions connected in series, the inductance of which changes with the electron temperature of the absorber.
    We capacitively shunt the SNS junctions by a $134\,\mathrm{pF}$ capacitor $C_{\mathrm{s}}$, which in combination with the Josephson inductance of the junctions forms an $LC$ oscillator.
    The absorber and thermometer are shorted to ground through a superconducting lead between them, which isolates them electrically while maintaining the thermal coupling.

When absorbing microwave photons, the incoming energy excites quasiparticles in the nanowire absorber and results in a new quasiequilibrium temperature on a timescale of nanoseconds.
This temperature rise decreases the critical current and increases the Josephson inductance of the short junctions on the thermometer side, which induces a shift of the resonance frequency of the $LC$ oscillator.
This shift is read out by reflecting a probe tone of frequency $f_{\mathrm{p}}$ and power $P_{\mathrm{p}}$ off of the gate capacitor $C_{\mathrm{g}}$, and by observing changes in the reflection coefficient $\Gamma$ near the resonance frequency.

    The device is cooled down to $20\,\mathrm{mK}$ in a dilution refrigerator.
    In both the bolometric and calorimetric operating modes, a continuous probe tone of frequency $f_{\mathrm{p}} \approx 600\,\mathrm{MHz}$ is reflected off of the device, and subsequently filtered, amplified, and digitized at room temperature in a heterodyne configuration (see Extended Data Fig.~\ref{fig:full_exp_setup} and Methods for the full experimental setup).
    An example of the recorded time-domain signal is shown in Fig.~\ref{fig:operating_principle}c, where a microwave pulse with length $t_{\mathrm{MW}} = 7\,\mathrm{ms}$, frequency $f_{\mathrm{MW}} = 8.40\,\mathrm{GHz}$, and power $P_{\mathrm{MW}} = -149.2\,\mathrm{dBm}$ at the chip input is applied, and the signal is averaged over $2^{13}$ repetitions.
    Here, a phase rotation has been applied to the in-phase ($I$) and quadrature-phase ($Q$) heterodyne components of the data such that the signal lies almost entirely in the $I$ component.
    The quasistatic responsivity $\delta V / \delta P_{\mathrm{MW}}$ is obtained from the difference between the signal during and before the pulse, whereas the thermal time constant $\tau$ is extracted from the exponentially rising edge of the signal at the arrival of the pulse (Methods).

\section*{Results and analysis}

    We first characterize the NEP of the sensor, which allows estimating the energy resolution (Methods).
    We apply an effectively constant ($t_{\mathrm{MW}} \gg \tau$) microwave input field, with the input power set to $P_{\mathrm{MW}} = -149.2\,{\rm dBm}$ to minimize the nonlinearity of the quasistatic response.
    The signal is averaged over $2^{13}$ repetitions to ensure a sufficient signal-to-noise ratio (SNR).

\begin{figure*}
    \centering
    \includegraphics[width=5.0in]{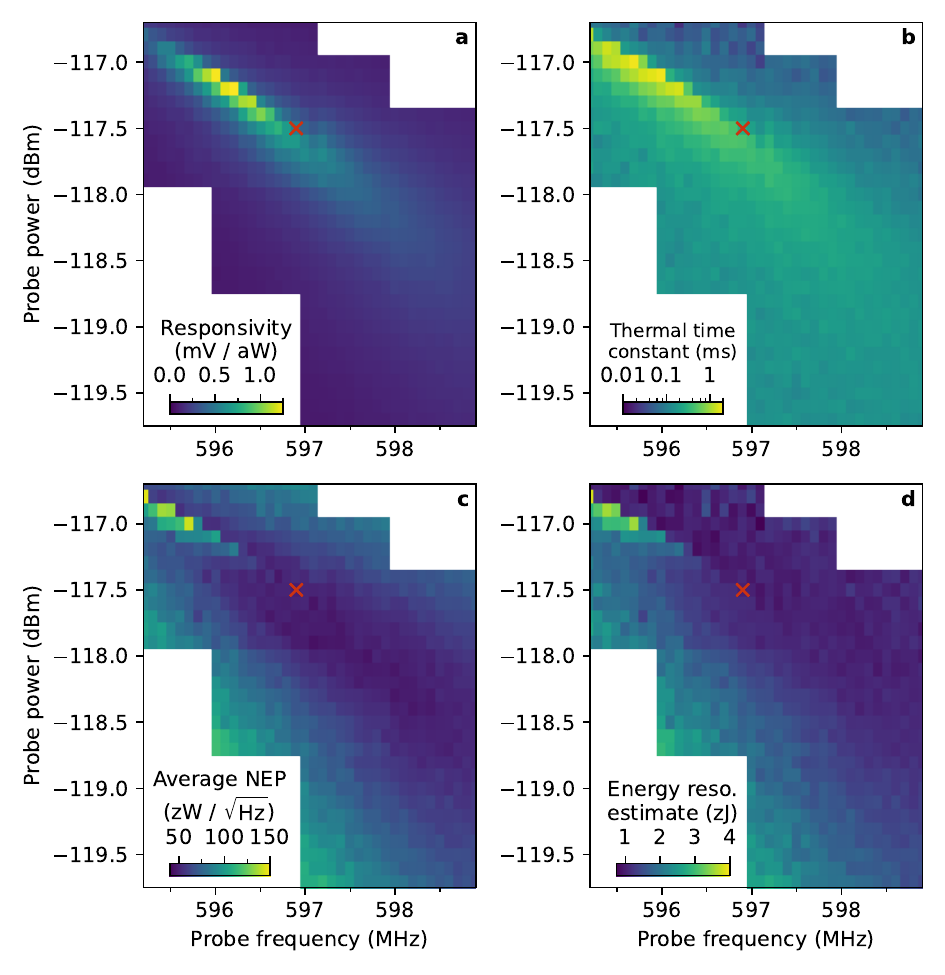}
    \caption{\textbf{Characterization results of the SNS sensor.}
    \textbf{a}, Quasistatic responsivity of the sensor as a function of probe frequency $f_{\mathrm{p}}$ and power $P_{\mathrm{p}}$.
    The responsivity is measured with a microwave signal power of $-149.2\,\mathrm{dBm}$.
    The red cross marks the operation point at $f_{\mathrm{p}} = 596.9\, \mathrm{MHz}$, $P_{\mathrm{p}} = -117.5\,\mathrm{dBm}$ chosen for the calorimetry.
    \textbf{b}, Rising-edge thermal time constant $\tau$ extracted from the time traces of the responsivity measurement by an exponential fit.
    \textbf{c}, Noise equivalent power (NEP) of the sensor averaged over $300\,\mathrm{Hz}$--$1\,\mathrm{kHz}$, as a function of probe frequency and power.
    \textbf{d}, Theoretical estimate for the FWHM energy resolution computed from the NEP data.
    See main text for more details.
    }
    \label{fig:NEP}
\end{figure*}

    The probe frequency $f_{\rm p}$ and power $P_{\rm p}$ are varied to find an optimal working point, see Fig.~\ref{fig:NEP}.
    The responsivity (Fig.~\ref{fig:NEP}a) is extracted from time-domain data similar to those shown Fig.~\ref{fig:operating_principle}c.
    The responsivity reaches its highest values slightly below the resonance frequency, where a change in the resonance frequency causes a large change in the transmission.
    Furthermore, the measured signal is proportional to the square root of the probe power, and thus increasing $P_{\mathrm{p}}$ increases the responsivity.
    However, a high probe power also exhibits Josephson non-linearity and electrothermal feedback\cite{Mccammon2005,Govenius2016}, owing to self-heating of the nanowire.
    This causes a shift in the resonance frequency, and the consequent shift in the maximum of the responsivity visible in Fig.~\ref{fig:NEP}a.
    Simultaneously, the shift induces a sharp increase in the thermal time constant, shown in Fig.~\ref{fig:NEP}b, even with a fixed microwave power $P_{\mathrm{MW}}$.
    The optimal probe parameters for the calorimetry are thus a trade-off between the responsivity and the time constant.
    At excessively high probe powers, the feedback causes bistability in the temperature of the nanowire, which can be utilized for highly sensitive threshold detection\cite{Govenius2016}, at the cost of significant dead time between measurements.
    However, we avoid this bistable region of the probe parameters, in order to operate our device as a calorimeter where the signal is proportional to the energy of the absorbed pulse, as opposed to a binary-valued click detector.

    In order to obtain the NEP, which is defined as the square root of the noise power spectral density (PSD) in the probe signal in units of the input microwave power, we additionally measure the noise PSD of the probe signal at each $f_{\mathrm{p}}$ and $P_{\mathrm{p}}$ (see Methods).
    Figure~\ref{fig:NEP}c shows the resulting NEP averaged over the frequency range $300\,\mathrm{Hz}$--$1 \, \mathrm{kHz}$.
    An NEP below $50\,{\rm zW}/ \sqrt{{\rm Hz}}$ is reached between probe powers of $P_{\mathrm{p}} = -118.5$ and $-117.5\,\mathrm{dBm}$.
    This is on par with previous record results achieved without a parametric amplifier at the millikelvin stage\cite{Roope_JPA}.

    From the measured frequency-dependent NEP, we estimate the corresponding calorimetric energy resolution of the sensor using equation~\eqref{eq:energy_reso_from_nep} of Methods and show the results in Fig.~\ref{fig:NEP}d.
    Note that the finest energy resolution does not exactly coincide with the lowest average NEP, owing to the fact that the energy resolution is integrated from zero up to $31.25\,\mathrm{MHz}$, the maximum sampling frequency we used to measure the noise spectrum, in contrast to the  $700\,\mathrm{Hz}$ averaging range used for the NEP.
    The finest estimated energy resolution is $0.74\, \mathrm{zJ}$ at $P_{\mathrm{p}} = -117.0\,\mathrm{dBm}$, $f_{\mathrm{p}} = 597.1\,\mathrm{MHz}$.
    However, at powers above $-117.5\,\mathrm{dBm}$, the bistability may be significant, and hence we choose the operation point with $P_{\mathrm{p}} = -117.5$, where $f_\mathrm{p} = 596.9\,\mathrm{MHz}$ yields the lowest NEP.
    At these probe parameters, we find $\tau = 260\,\mathrm{\mmu s}$, average $\mathrm{NEP} = 49\,\mathrm{zW / \sqrt{Hz}}$, and an estimated energy resolution of $1.03\,\mathrm{zJ}$.

    Having found this promising operation point, we move to calorimetric measurements in order to measure the energy resolution instead of merely extracting it from the NEP.
    Here, we measure probe signal traces with no ensemble-averaging, while sending short, $1\,\mathrm{\mmu s} \ll \tau$ pulses at $8.40\,\mathrm{GHz}$ to the calorimeter input, with pulse energies ranging between $0.95\,\mathrm{zJ}$ and $3.8\,\mathrm{zJ}$ ($\approx 170$--$680$ photons).
    The pulse energies are accurately known since the input line attenuation has been carefully calibrated (Methods).

\begin{figure}
    \centering
    \includegraphics[
    width=3.5in
    ]{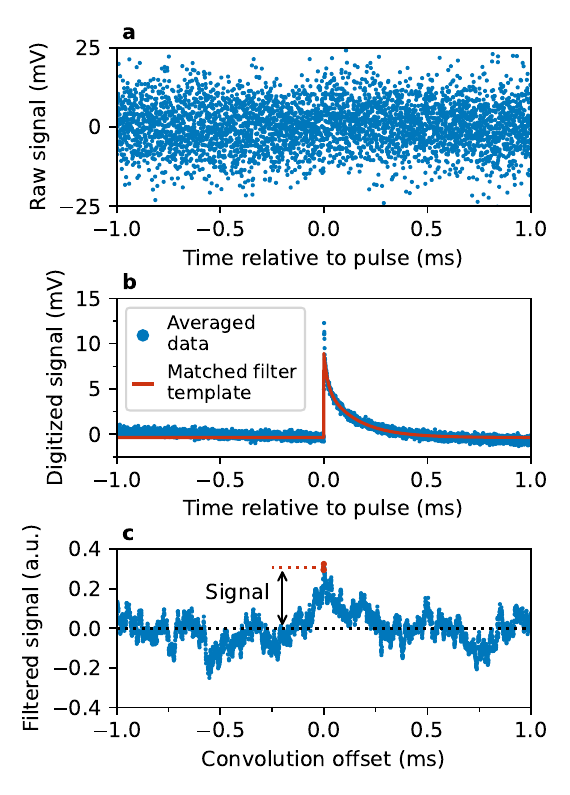}
    \caption{\textbf{Matched-filtering procedure.}
    \textbf{a}, Raw digitized time trace of the probe signal in a single-shot measurement, with a $1 \, \mathrm{\mmu s}$ calorimeter input pulse having a mean energy of $1.19\,\mathrm{zJ}$ applied at $t = 0$.
    \textbf{b}, As \textbf{a} but for a $1\, \mathrm{\mmu s}$ input pulse having energy of approximately $3.8\,\mathrm{zJ}$ averaged over $1000$ repetitions (blue dots). A fit to a model that is a sum of two exponential decays is also shown (red curve).
    The model is used as the template of the matched filter (Methods).
    \textbf{c}, Data of \textbf{a} after application of the matched filter according to the fitted template in \textbf{b}, as a function of the convolution offset between the template and the data.
    The calorimetric signal $\bar S$ is obtained as the value of the filtered signal averaged over a $1\, \mathrm{\mmu s}$ range around zero offset, highlighted in red.
    }
    \label{fig:matched_filtering}
\end{figure}

    Such low pulse energies result in an extremely low SNR in the raw probe signal traces, as shown in Fig.~\ref{fig:matched_filtering}a.
    Consequently, we employ a matched filter to significantly improve the SNR.
    To extract a template for the matched filter, we average 1000 pulses with $3.8\,\mathrm{zJ}$ of energy and fit a model for the expected temporal envelope of the signal (see Fig.~\ref{fig:matched_filtering}b and Methods).
    Figure~\ref{fig:matched_filtering}c shows the output of the matched filter, as a function of the offset of the convolution between the transfer function of the filter and the input data.
    We average over a $1\,\mathrm{\mmu s}$ window around the zero convolution offset to obtain the calorimetric signal $\bar S$.
    For comparison, we also carry out a simple averaging of the non-matched-filtered data over varying time windows.
    We find that for the $0.95\,\mathrm{zJ}$ pulse, the SNR from matched filtering is approximately 30\% higher than the highest SNR with simple averaging, which is achieved with a $27\,\mathrm{\mmu s}$ window.

\begin{figure}
    \centering
    \includegraphics[
    width=3.5in
    ]{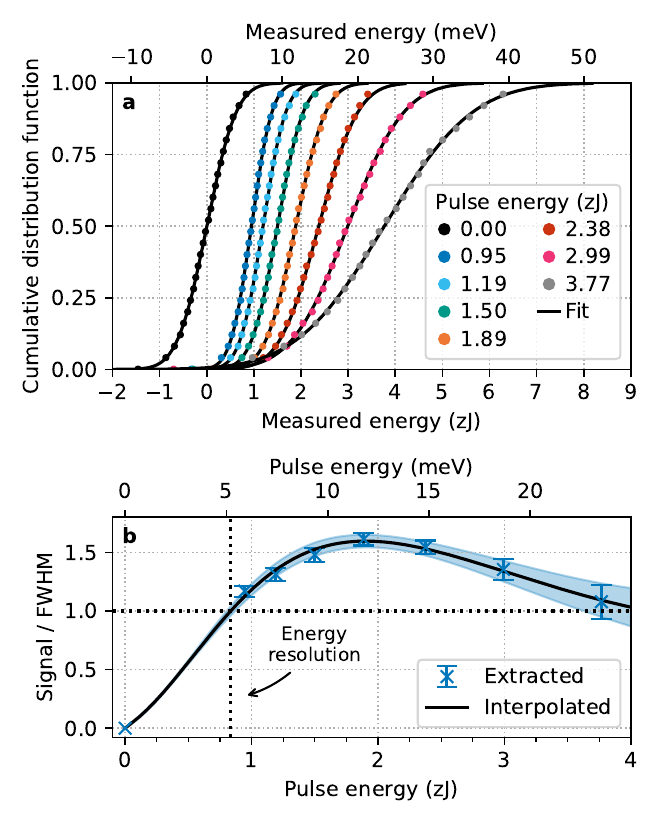}
    \caption{
        \textbf{Zeptojoule calorimetry.}
        \textbf{a}, Empirical cumulative distribution function (CDF) of the calorimetric signal converted to energy units as a function of pulse energy (dots).
        The CDF is calculated from 1000 time traces after matched filtering.
        Only one in every 40 points for each energy is shown for clarity.
        The solid lines exhibit fits of the error function to the experimental data.
        \textbf{b}, Mean values of the distribution of \textbf{a} divided by their full widths at half maxima (FWHMs) as functions of the pulse energy (crosses), with intermediate values (line) obtained by interpolating the fit parameters extracted from \textbf{a}.
        The energy resolution is the energy at which the mean signal is equal to the FWHM, indicated by the vertical dotted line.
        The error bars and shaded regions denote one-standard-deviation confidence intervals for the measured and interpolated data, respectively.
    }
    \label{fig:calorimetry}
\end{figure}

    Next, we construct empirical cumulative distribution functions (CDFs) of the calorimetric signals $\bar S$ from $1000$ traces for each pulse energy.
    We find that the noise in the signal is well approximated by a normal distribution, and we thus obtain a good agreement with fits of the CDFs to the error function, as shown in Fig.~\ref{fig:calorimetry}a.
    From these fits, we extract the means $\mu_{\bar S}$ and standard deviations $\sigma_{\bar S}$ of the distributions for each pulse energy.
    We convert these to the mean $\mu_E$ and standard deviation $\sigma_E$ of the energy measured by the calorimeter using the calibrated energies of the pulses reaching the chip, and a model for the dependence of the signal on the pulse energy (see Methods).

    With large pulse energies, the resonance frequency of the tank circuit moves far away from the probe frequency, and thus the probe signal effectively saturates.
    This is reflected in the width of the distribution of the measured energy increasing significantly in Fig.~\ref{fig:calorimetry}a.
    Note that fluctuations in the resonance frequency, such as those owing to thermal fluctuations in the thermometer, or the Poissonian distribution of photons in the input pulse\cite{keranen2024}, result in a distribution for $\bar S$ that is distorted by the Lorentzian dependence of the signal on the resonance frequency.
    This effect should be especially pronounced for low pulse energies and $f_{\mathrm{p}}$ chosen close to the resonance, as is our case.
    Yet, we find the signal to very closely follow a normal distribution.
    Furthermore, at high powers, fluctuations in the signal $\bar S$ owing to fluctuations in the resonance frequency should reduce as a result of the aforementioned saturation.
    This is the opposite of what we find for $\sigma_{\bar S}$ before conversion to energy (see Methods).
    This suggests that the noise in the output signal is primarily arising from the amplification chain, and our experiment is therefore not limited by the sensor itself.
    This observation is consistent with the results of Ref.~\cite{Roope_JPA}, where placing a quantum-limited amplifier directly at the probe output at the millikelvin stage of the cryostat significantly improved the NEP and the estimated energy resolution.

    We determine the energy resolution of our calorimeter by the input energy that yields a signal-to-FWHM ratio of unity, i.e., resolving power of unity, as shown in Fig.~\ref{fig:calorimetry}b.
    Although no data with a pulse energy below $E_{\mathrm{MW}} = 0.95\,\mathrm{zJ}$ is available, the signal-to-FWHM ratio with $(0.95 \pm 0.02)\, \mathrm{zJ}$ pulse energy is $1.17 \pm 0.05 > 1$, which thus implies that the energy resolution of the calorimeter is finer than $(0.95 \pm 0.02)\, \mathrm{zJ}$.
    This corresponds to approximately $171\pm4$ photons at the microwave signal frequency of $8.40\, \mathrm{GHz}$.
    The uncertainty of $0.02 \, \mathrm{zJ}$ arises from an uncertainty of $0.1\,\mathrm{dB}$ in the calibration of the input line attenuation (see Methods).
    By interpolating $\sigma_E$ (see Methods), we estimate the actual energy resolution of the calorimeter to be $(0.83\pm 0.04)\,\mathrm{zJ}$, corresponding to approximately $150\pm 7$ photons.
    This agrees reasonably well with the energy resolution of $1.03\,\mathrm{zJ}$ predicted by the NEP.
    
    Owing to the large shift of the resonance, the signal-to-FWHM ratio reaches a maximum near $E_\textrm{MW} = 2\,\mathrm{zJ}$ and drops below unity at higher energies, signifying a dynamic range of only a few $\mathrm{zJ}$.
    This could be mitigated, for example, by utilizing a frequency comb or simply a second probe tone at a frequency near the shifted resonance, but in this work, we have focused on optimizing the energy resolution instead of the dynamic range.

\section*{Conclusions}

In conclusion, we have demonstrated the operation of a radiation sensor based on SNS junctions in a calorimetric mode.
At our chosen operation point, the thermal time constant $\tau = 260\,\mathrm{\mmu s}$ and $\mathrm{NEP} = 49\,\mathrm{zW / \sqrt{Hz}}$ of the device are on par with the state-of-the-art values for ultrasensitive bolometers.
We have measured the FWHM energy resolution to be finer than $0.95\,\mathrm{zJ} = 5.9\,\mathrm{meV}$ when detecting $8.4\,\mathrm{GHz}$ microwave signals.
To the best of our knowledge, this is the finest energy resolution reported for any calorimeter.
By interpolating our data, we further estimate that the energy resolution may, in fact, be as fine as $0.83\,\mathrm{zJ} = 5.2\,\mathrm{meV}$.

An important development in reaching this result was the use of a matched filter for the SNS sensor, which improved the SNR by more than 30\% compared to simple averaging in the time domain.
%
%
%
Our results indicate that the dominating source of noise in the measured signal is the amplification chain after our device, as has been reported previously for similar devices.
This suggests that the energy resolution could be significantly improved by using a quantum-limited parametric amplifier at the millikelvin stage of the cryostat.
We also expect that using similar data analysis methods as presented here, SNS sensors based on absorbers with lower heat capacity, such as graphene, could further substantially improve the energy resolution.
In the future, we aim to pursue this appealing experiment together with demonstrating the case where the arrival time of the signal to be detected is unknown, which is relevant for some practical applications, such as the detection of cosmic radiation.

A radiation sensor with sub-zeptojoule energy resolution is of great interest for a broad range of applications, including astrophysics and cosmology, quantum technology, and quantum thermometry.
Our work lays the foundation for such applications and eventually for calorimetry down to the single-microwave-photon level. Immediate advantages may be obtained by adapting the matched-filtering scheme of this work to the calorimetric readout of superconducting qubits\cite{Gunyho2024}.

\section*{Methods}

\setcounter{figure}{0}
\renewcommand{\figurename}{\textbf{Extended Data Fig.}}
\renewcommand{\theHfigure}{Extended.\thefigure}  

\setcounter{table}{0}
\renewcommand{\tablename}{\textbf{Extended Data Table}}
\renewcommand{\theHtable}{Extended.\thetable}


\subsection*{Full experimental setup}

\begin{figure*}
    \centering
    \includegraphics[width=5.0in]{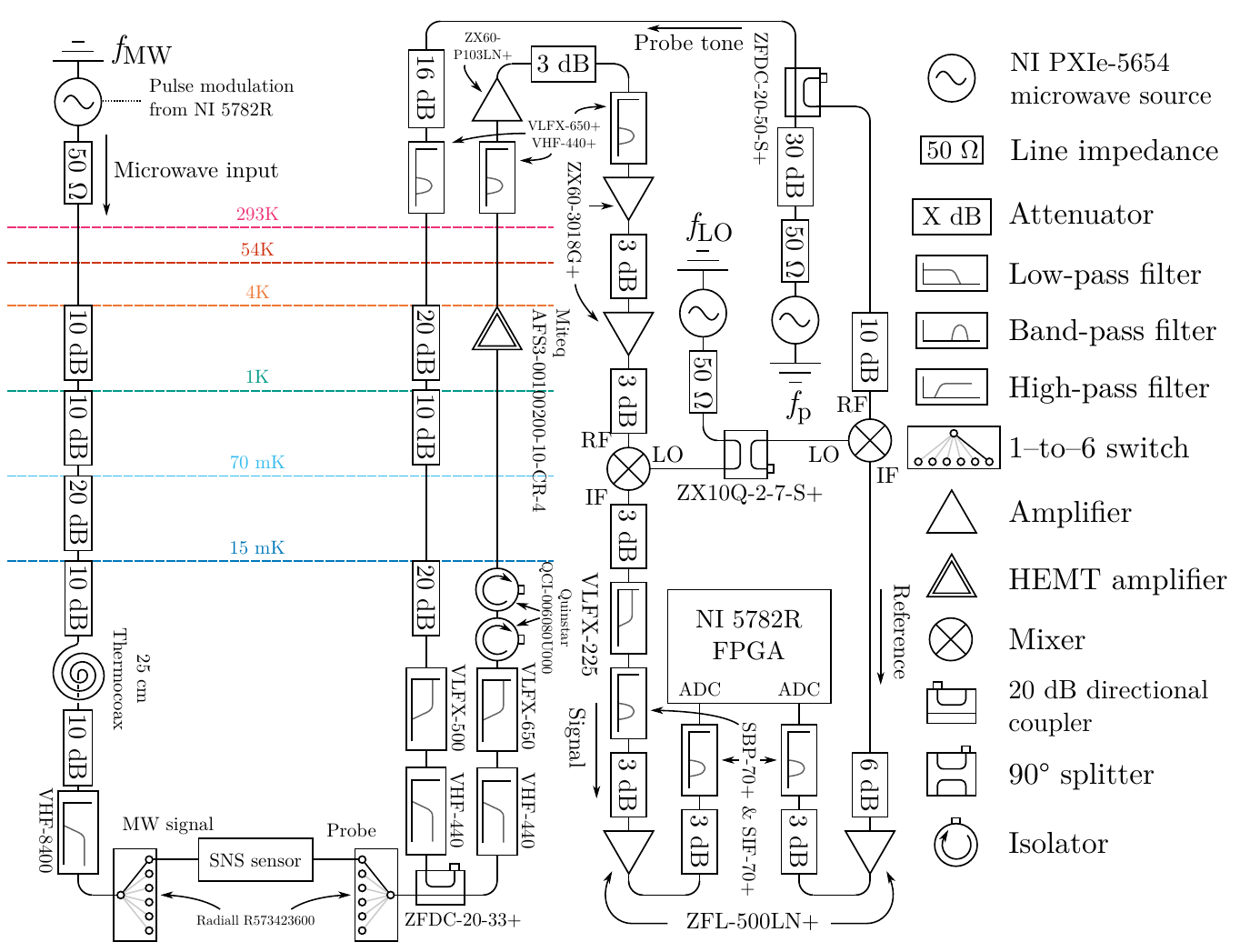}
    \caption{
        \textbf{Full experimental setup.}
        Labels below components denote Mini-Circuits model numbers.
        The band-pass filters consist of a low-pass filter and high-pass filter in series.
        The NI 5782R FPGA board handles triggering the microwave pulse via pulse modulation, as well as analog-to-digital conversion (ADC) and data acquisition.
        A control computer (not shown) is used to initiate each measurement and collect the data.
    }
    \label{fig:full_exp_setup}
\end{figure*}

A detailed depiction of the experimental setup is shown in Extended Data Fig.~\ref{fig:full_exp_setup}.
A NI 5782R transceiver module sends digital triggers to apply pulse modulation to a PXIe-5654 microwave source to generate the microwave pulses.

The continuous probe tone is generated by another NI PXIe-5654 microwave source.
The signal is split using a directional coupler into the probe signal entering the cryostat and a reference signal.
The probe tone is attenuated and filtered, and reflected off of the SNS sensor via a directional coupler.
After being reflected, the signal is amplified by a high-electron-mobility-transistor (HEMT) amplifier at the $4\,\mathrm{K}$ stage of the cryostat, and then by room temperature low-noise amplifiers.
The probe signal is demodulated from radio frequency (RF) by mixing it with a local oscillator (LO) signal, which is always detuned from $f_{\mathrm{p}}$ by the fixed intermediate frequency (IF) $70.3125\,\mathrm{MHz}$.
After demodulation, the signal is further filtered to remove sidebands from the mixing, and amplified more at the IF frequency.

The probe signal is digitized by the NI 5782R at a sampling rate of $250\,\mathrm{MSa/s}$ and digitally demodulated from the IF frequency to DC, yielding the heterodyne in-phase ($I$) and quadrature-phase ($Q$) components.
Finally, the signal is averaged and decimated over $2^7$ adjacent samples for a digital time step of $512\,\mathrm{ns}$.
The reference signal is directly demodulated and digitized, and it is used as a phase reference for the digitized probe tone.

Attenuators are placed between amplifiers in order to avoid standing waves from reflections between the components.
The 1--to--6 switches in Extended Data Fig.~\ref{fig:full_exp_setup} are used to switch between the sample used in our experiment and other samples unrelated to this work.

    The total input line attenuation has been calibrated following the procedure introduced in Ref.~\cite{Girard2023}.
    In a separate thermal cycle of the cryostat, we replace the SNS sensor used in the main experiment with a bolometer that can be heated both with microwave power and DC current.
    We apply a known DC power to the nanowire using a four-wire configuration, and by varying the RF powers applied at room temperature and comparing the resulting frequency shift, we find the correspondence between the room-temperature RF power and the power at the chip.
    We find the total input line attenuation to be $119.24\pm 0.1\,\mathrm{dB}$.
    Note that this is the total attenuation up to the input of the sample holder containing the chip used in the main experiment.
    Thus the reported energy resolution considers our sensor to be a black box at the end of a $50\,\mathrm{\Omega}$ transmission line, and it includes possible reflections owing to impedance mismatch at the chip input which degrade the measured energy resolution.
    We consider the change in line attenuation between the thermal cycles of the cryostat to be negligible.

\subsection*{Model for bolometric transmission}

    In the bolometric linear mode of operation, the steady-state reflection coefficient of the gate capacitor at a given probe frequency $f_{\mathrm{p}}$ follows a Lorentzian line shape~\cite{Chen2022Scattering2}:
    \begin{equation}
        \Gamma \propto
        1 - \frac{\ee^{\ii\varphi}\gamma_{\mathrm{c}}}{\gamma/2 + 2\pi \ii \left(f_{\mathrm{r}} - f_{\mathrm{p}}\right)}
        ,
    \end{equation}
    where $f_{\mathrm{r}}$ is the resonance frequency of the $LC$ tank circuit of the sensor, $\gamma_{\mathrm{c}}$ and $\gamma$ are the external and total energy decay rates, respectively, and $\varphi$ is a parameter corresponding to the asymmetry of the resonance owing to impedance mismatch between the chip and the coaxial cable at the probe input.
    The resonance frequency shifts approximately linearly with the power $P_{\mathrm{MW}}$ of the microwave tone~\cite{keranen2024} (or equivalently, the pulse energy $E_{\mathrm{MW}} = P_{\mathrm{MW}} \, t_{\mathrm{MW}}$):
    \begin{equation}
        f_{\mathrm{r}}
        =
        f_{\mathrm{r}, 0} - \alpha P_{\mathrm{MW}}
        ,
    \end{equation}
    where $f_{\mathrm{r},0}$ is the resonance frequency with no microwave pulse, and $\alpha$ is a coefficient with units of $\mathrm{Hz} / \mathrm{W}$ corresponding to the conversion of $P_{\mathrm{MW}}$ to heat in the absorber, and subsequently to a change in the Josephson inductance of the nanowire.
    In general, $\alpha$ depends on several factors, including the absorber geometry and material, and impedance matching of the absorber to the input coaxial line.

    The difference in the complex heterodyne transmitted signal $I + \ii Q$ with a given input power is thus proportional to
    \begin{multline}
        \label{eq:lorentzian_diff}
        \Gamma(P = P_{\mathrm{MW}})
        -
        \Gamma(P = 0)
        \\
        \propto
        \left(
            \frac{1}{\gamma/2 + 2\pi \ii \left(f_{\mathrm{r,0}} - f_{\mathrm{p}}\right)}
            -
            \frac{1}{\gamma/2 + 2\pi \ii \left(f_{\mathrm{r,0}} - \alpha P_{\mathrm{MW}} - f_{\mathrm{p}}\right)}
        \right)
        \ee^{\ii\varphi}\gamma_{\mathrm{c}}
        .
    \end{multline}
    The relative change in the digitized voltage $\Delta V$ is obtained by applying a rotation to cancel out $\varphi$, multiplying by the total gain of the amplification chain $G$, and discarding the imaginary part as
    \begin{align}
        \label{eq:lorentzian_diff_re}
        \nonumber
        \Delta V
        &
        =
        G
        \gamma_{\mathrm{c}}
        \operatorname{Re}\left[
            \frac{1}{\gamma/2 + 2\pi \ii \left(f_{\mathrm{r,0}} - f_{\mathrm{p}}\right)}
            -
            \frac{1}{\gamma/2 + 2\pi \ii \left(f_{\mathrm{r,0}} - \alpha P_{\mathrm{MW}} - f_{\mathrm{p}}\right)}
        \right]
        \\
        &=
        G
        \frac{\gamma \gamma_{\mathrm{c}}}{2}
        \left(
            \frac{1}{(\gamma/2)^2 + \left[2\pi \left(f_{\mathrm{r},0} - f_{\mathrm{p}}\right) \right]^2}
            -
            \frac{1}{(\gamma/2)^2 + \left[2\pi \left(f_{\mathrm{r},0} - \alpha P_{\mathrm{MW}} - f_{\mathrm{p}}\right) \right]^2}
        \right)
        .
    \end{align}
    With large values of $P_{{\mathrm{MW}}}$, this expression becomes effectively constant, and hence the energy resolution of the sensor approaches zero.
    However, it may still be utilized as a binary detector, indicating the presence or absence of a microwave signal.

\subsection*{Estimation of the noise power spectral density}

    The one-sided noise power spectral density (PSD) $S_{\mathrm{n}}$ of the measured voltage is used both in calculating the NEP and in the matched-filtering procedure.
    We estimate the noise PSD by averaging periodograms according to Bartlett's method\cite{proakis1996} for data with no pulse applied.

\begin{figure*}
    \centering
    \includegraphics[width=5in]{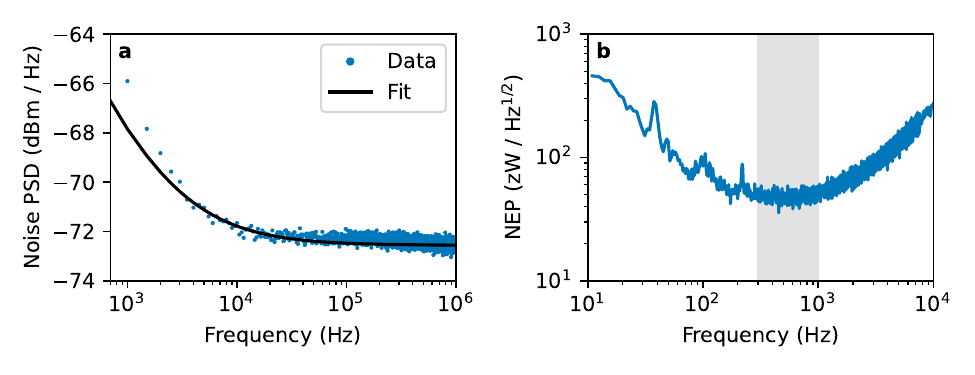}
    \caption{
        \textbf{Noise power spectral density (PSD) and noise equivalent power (NEP) at the operation point}
        \textbf{a}, Noise power spectral density as a function of noise frequency (points), and a fit to the heuristic model $A/f + B$ (solid line).
        \textbf{b}, Noise equivalent power as a function of noise frequency.
        The shaded region indicates the range over which the NEP is averaged to obtain the data in Fig.~\ref{fig:NEP}c.
    }
    \label{fig:noise_psd_and_nep}
\end{figure*}

    The PSD used in the matched filtering is obtained by fitting the estimated PSD to a simple heuristic model $\tilde S_{\mathrm{n}}(f) = A/f + B$, which is a sum of $1/f$ and white noise.
    Using this model instead of the estimated PSD directly yields a better signal-to-noise ratio, since the estimate is noisy (see Extended Data Fig.~\ref{fig:noise_psd_and_nep}a).
    The measured spectrum does not completely agree with this model, which we attribute to the analog filtering and amplification in the output chain.
    Nevertheless, we find that using the model in the matched-filtering procedure improves the energy resolution by approximately 4\% compared with using the raw noise PSD.
    Note that one may freely choose the noise model in the matched-filtering procedure without loss of scientific soundness.

\subsection*{Obtaining the noise equivalent power}

In order to obtain the NEP from the experimental data in the bolometric mode, we record time-domain traces of the transmitted signal similar to the one shown in Fig.~\ref{fig:operating_principle}c for different values of $f_{\mathrm{p}}$ and $P_{\mathrm{p}}$.
We offset the time axis such that $t = 0$ is at the arrival time of the pulse, and fit these traces to the model $\Delta V \cdot \left(1 -  e^{-t/\tau}\right)$ for $t > 0$ in order to extract the relative-voltage signal $\Delta V$ and thermal time constant $\tau$.

With $f_{\mathrm{p}}$ close to $f_{\mathrm{r}, 0}$ and $\alpha P_{\mathrm{MW}} \ll \gamma$, we have an approximately linear dependence of $\Delta V$ on $P_{\mathrm{MW}}$, and hence we define the quasistatic responsivity as $\delta V / \delta P_{\mathrm{MW}} = \Delta V / P_{\mathrm{MW}}$, where the word \emph{quasistatic} refers to the fact that this definition assumes that $P_{\mathrm{MW}}$ is constant in time.
In order to determine how the responsivity varies with temporal changes in $P_{\mathrm{MW}}$ occurring at the noise frequency $f_{\mathrm{n}}$, the responsivity may be measured while modulating $P_{\mathrm{MW}}$ at $f_{\mathrm{n}}$.
However, we capture the frequency dependence of the responsivity by~\cite{Mccammon2005, Roope_JPA, Devisser2014} $\delta V / \delta P_{\mathrm{MW}} \left[1 + \left(2\pi f_{\mathrm{n}}\tau\right)^2\right]^{-1/2}$, which takes into account the decrease in responsivity owing to the thermal time constant $\tau$.
Note that this procedure is justified by our empirical observation in Fig.~\ref{fig:operating_principle}c that the rise of the sensor signal for $P_{\mathrm{MW}}$ suddenly turned on is accurately exponential.

The NEP is thus given by
\begin{align}
    \operatorname{NEP}\left(f_{\mathrm{n}}\right)
    &
    =
    \sqrt{S_{\mathrm{n}}\left(f_{\mathrm{n}}\right)}
    \left(\delta V / \delta P_{\mathrm{MW}}\right)^{-1}
    \sqrt{1 + \left(2\pi f_{\mathrm{n}}\tau\right)^2}
    ,
\end{align}
where $S_{\mathrm{n}}(f_{\mathrm{n}})$ is the noise PSD of the measured relative voltage $\Delta V$ at the frequency $f_{\mathrm{n}}$.
The units of the NEP are ${\rm zW}/ \sqrt{{\rm Hz}}$ since it equals the square root of the noise power spectral density of the measured signal in units of input power to the bolometer.
The NEP obtained at the operation point used for the calorimetry is shown in Extended Data Fig.~\ref{fig:noise_psd_and_nep}b.

For the probe parameters considered in Fig.~\ref{fig:NEP}, the NEP is lowest around the range of $0.3$--$1\,\mathrm{kHz}$, indicated by the shaded region in Extended Data Fig.~\ref{fig:noise_psd_and_nep}b.
In this frequency range, the effect of the $1/f$ noise is reduced, but the insensitivity owing to finite $\tau$ is not yet significant.
This implies that the highest sensitivity in the bolometric mode can be achieved by modulating the input power $P_{\mathrm{MW}}$ at a frequency within this range, for example, by using a shutter, a multiplexer that switches between two bolometers, or by employing lock-in detection.
Thus we choose to average over this range for the data shown in Fig.~\ref{fig:NEP}c.

The NEP also allows us to estimate the energy resolution, through the relation~\cite{Mccammon2005}
\begin{align}
    \label{eq:energy_reso_from_nep}
    \Delta E_{\mathrm{FWHM}}
    &
    =
    2\sqrt{2 \ln 2}
    \left(
    \int_0^{f_\textrm{max}} \frac{4}{\mathrm{NEP}^2(f)} \mathrm{d} f
    \right)^{-1/2}
    ,
\end{align}
where the factor $2\sqrt{2 \ln 2}$ arises from the used full width at half maximum (FWHM) of a normally distributed signal (discussed below) and $f_\textrm{max}$ is a frequency, up to which the sensor is employed.
In our previous work\cite{Roope_JPA}, we have used the thermal cut-off $1/(2\pi\tau)$ for the upper bound of the integral, but we find that in our case, this gives an unnecessarily pessimistic estimate.
This equation is used to obtain the data shown in Fig.~\ref{fig:NEP}d.

\subsection*{Single-shot data processing using matched filtering}

    Here, we discuss the data processing employed for the data in the calorimetry experiments where no ensemble-averaging is carried out.
    The phase of each digitized and down-converted heterodyne trace relative to each other is normalized using a phase reference (see Extended Data Fig.~\ref{fig:full_exp_setup}), and the baseline is removed by subtracting the median of the $I$ and $Q$ components of each trace for the section before the pulse.
    The baseline-removed traces are rotated in the $IQ$-plane by a global phase angle $\tilde \varphi$ such that projecting the rotated signal to the $I$ axis yields the finest energy resolution.
    An example of such a digitized trace after this preprocessing step is shown in Fig.~\ref{fig:matched_filtering}a.

    After the preprocessing, a matched filter\cite{wainstein1962,Mccammon2005} is applied to the projected in-phase signal, which maximizes the signal-to-noise ratio.
    For the matched filter, the filtered signal is given by the following convolution:
    \begin{equation}
        S_k = \operatorname{DFT}^{-1} \left[
            \frac{
                \operatorname{DFT} \left[ V_k \right]_j
                \cdot
                \overline{\operatorname{DFT} \left[ K_k \right]}_j
            }{
                S_{\mathrm{n},j}
            }
        \right]_k
        ,
    \end{equation}
    where
    $V_k$ is the digitized signal at the sample index $k$,
    $K_k$ is the \emph{template} of the matched filter (discussed below),
    $\overline{z}$ denotes the complex conjugate of $z$,
    $S_{\mathrm{n},j} = S_{\mathrm{n}}\left(f_{\mathrm{n},j}\right)$ is the noise PSD at the frequency bin with index $j$,
    and
    $\operatorname{DFT}$  and $\operatorname{DFT}^{-1}$ denote the discrete Fourier transform and its inverse, respectively.

    The values of the signal after the matched filtering, $S_k$, indicate how well the raw signal and template correlate at a given convolution offset, with additional weighting from the noise PSD emphasizing frequency components that are less noisy.
    If the pulse arrival time is known and the template has a duration equal to the window of the digitized signal, the matched filter yields a peak at zero convolution offset.
    The final calorimetric signal $\bar S$ is obtained by taking the mean of the filtered signal $S_k$ over a $1.024\,\mathrm{\mmu s}$ window centered at zero convolution offset, as shown in Fig.~\ref{fig:matched_filtering}c.

    The template of the matched filter is the expected time-domain shape of the signal, which is assumed to be directly proportional to the energy of the pulse.
    For the short pulses used in the calorimetry, we model the temporal dependence of the projected signal as a sum of two exponentially decaying processes as
    \begin{equation}
        \label{eq:template}
        K(t) = \begin{cases}
            0 & t < 0, \\
            (a_1 + a_2) \cdot
            t / t_{\mathrm{MW}}
              & 0 \leq t < t_{\mathrm{MW}}, \\
              a_1 \exp\left[-(t - t_{\mathrm{MW}}) / \tau_1\right]
              +
              a_2 \exp\left[-(t - t_{\mathrm{MW}}) / \tau_2\right]
              & t_{\mathrm{MW}} \leq t,
        \end{cases}
    \end{equation}
    where the arrival time of the pulse is $t = 0$ and $t_{\mathrm{MW}} = 1\,\mathrm{\mmu s}$ is the length of the pulse.
    The amplitude $a_1$ and time constant $\tau_1$ correspond to a fast decay of heat from the electrons to an intermediate heat bath, and $a_2$ and $\tau_2$ to a slow decay from the intermediate bath to the phonon bath of the chip substrate through a weak thermal link.
    The exact physical origin of this double-exponential behavior is unknown, but such behavior has been reported in earlier experiments with similar devices\cite{Govenius2016,Gasparinetti_bolometer}.
    Here, we assume a zero baseline, and that the length of the input pulse $t_{\mathrm{MW}}$ is much shorter than the time constants $\tau_1$ and $\tau_2$, so that the signal rises effectively linearly during the pulse.
    We extract the template by ensemble-averaging 1000 pulses with a relatively high pulse energy of $3.8\,\mathrm{zJ}$ and fitting the data to equation~\eqref{eq:template}.
    From the fit, we obtain $\tau_1 \approx 18\,\mathrm{\mmu s}$, $\tau_2 \approx 150\,\mathrm{\mmu s}$, and $a_2 / a_1 \approx 1.5$.
    The template with these parameters is shown in Fig.~\ref{fig:matched_filtering}b.

    With high powers, the heterodyne signal moves along a curve in the $IQ$ plane, as predicted by equation~\eqref{eq:lorentzian_diff}.
    Thus the linear relationship between the pulse energy and the projected signal $\Delta V$ breaks down, and the effectiveness of the matched filter is reduced.
    While we observe some deviation from the linear behavior in our data, we find that this effect does not significantly affect the filtering procedure for the pulse energies considered here.

\subsection*{Conversion of filtered signal to energy units and the energy resolution calculation}

    The calorimetric signal $\bar S$ extracted from the matched-filtering procedure has arbitrary units, since we have not calibrated the whole amplification chain from the sample to the analog-to-digital converter. 
    Here, we discuss the procedure used to convert the extracted signal and its standard deviation to units of energy for Fig.~\ref{fig:calorimetry}, and how that is subsequently used to obtain the energy resolution.

    For each pulse energy $E_{\mathrm{MW}}$, we collect $N = 1000$ traces and extract the corresponding signals $\bar S_j$, $j = 1, \dots, N$ as discussed above.
    We then compute the empirical cumulative distribution function (CDF) for each $E_{\mathrm{MW}}$, given by
    \begin{align}
        \operatorname{CDF}_{E_{\mathrm{MW}}}(s)
        &
        =
        \sum_{\bar S_j \leq s} \frac{1}{N}
        .
    \end{align}
    The noise in the signal closely follows a normal distribution, and therefore we model the CDF for a given pulse energy $E$ as
    \begin{align}
        \label{eq:cdf_model}
        \operatorname{CDF}_E(s)
        &=
        \frac{1}{2}
        +
        \frac{1}{2} \operatorname{erf}\left(\frac{s-\mu_{\bar S}(E)}{\sqrt{2} \sigma_{\bar S}(E)}\right)
        ,
    \end{align}
    where $\operatorname{erf}(\cdot)$ is the error function, and $\mu_{\bar S}$ and $\sigma_{\bar S}$ are the mean and standard deviation of the corresponding normal distribution, respectively.
    We extract $\mu_{\bar S}$ and $\sigma_{\bar S}$ by a least-square fit of the error function in equation~\eqref{eq:cdf_model} to the empirical CDF.

\begin{figure*}
    \centering
    \includegraphics[width=5in]{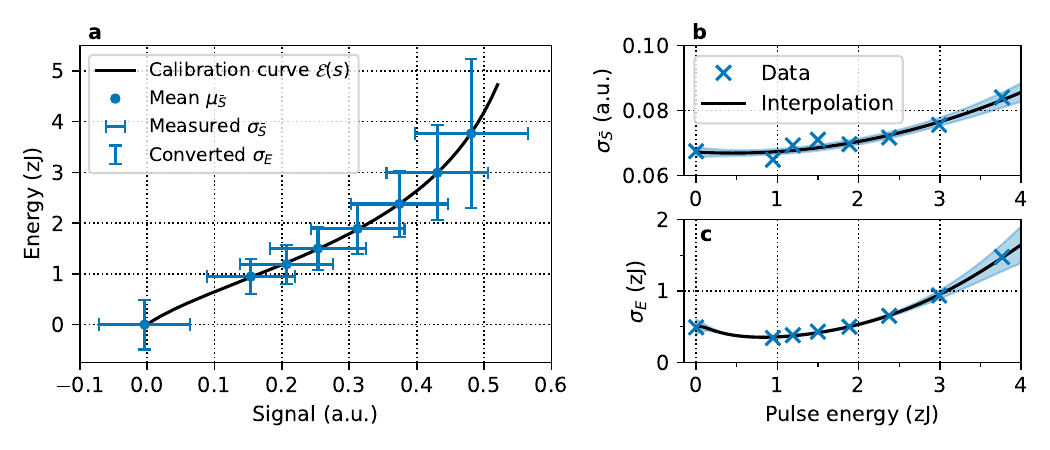}
    \caption{
        \textbf{Conversion between the measured calorimetric signal and the energy at the input of the sensor.}
        \textbf{a}, Mean energy in the input pulses of the calorimeter as a function of the mean value $\mu_{\bar S}$ of the normal distribution of the corresponding measured signal distribution (dots). Conversion of arbitrary signal level to energy units is obtained from the calibration curve $\mathcal{E}(s)$ (solid line).
        The horizontal error bars denote the standard deviation in signal units, and the vertical bars the corresponding uncertainty in energy, obtained by transforming the variance by $\mathcal{E}$ (see text).
        \textbf{b}, Standard deviation $\sigma_{\bar S}$ in signal units (crosses) as a function of the calibrated pulse energy.
        The solid line shows a quadratic polynomial that is used to interpolate the values of $\sigma_{\bar S}$.
        \textbf{c}, Standard deviation (crosses) and the interpolated curve (solid line) after transforming the data of \textbf{b} using $\mathcal{E}$.
        The shaded regions in panels \textbf{b} and \textbf{c} indicate the one-standard-deviation confidence intervals $\delta \sigma_s$ and $\delta \sigma_E$, respectively.
    }
    \label{fig:energy_conversion}
\end{figure*}

    Since we have calibrated the total attenuation at the input of the microwave absorber, the energy corresponding to each value of $\mu_{\bar S}$ is known, up to a constant relative error of $\pm 0.1 \, \mathrm{dBm}$, discussed below.
    Since $\bar S$ is proportional to $\delta V / \delta P_{\mathrm{MW}}$, its dependence on $E_{\mathrm{MW}}$ is also of the form of equation~\eqref{eq:lorentzian_diff_re} (with $P_{\mathrm{MW}} = E_{\mathrm{MW}} / t_{\mathrm{MW}}$ and $\alpha$ adjusted).
    We invert equation~\eqref{eq:lorentzian_diff_re} by solving for $E_{\mathrm{MW}}$ to obtain
    \begin{align}
        \mathcal{E}(s)
        &
        =
        \frac{t_{\mathrm{MW}}}{2\pi\alpha}
        \left(
            2\pi
            \left(
            f_{\mathrm{r},0}
            -
            f_{\mathrm{p}}
            \right)
            +
            \sqrt{
                \left[
                    \frac{1}{
                        (\gamma/2)^2
                        +
                        \left[2\pi\left(f_{\mathrm{r},0} - f_{\mathrm{p}}\right)\right]^2
                    }
                    -
                    \frac{s}{a}
                \right]^{-1}
                - 1
            }
        \right)
        ,
    \end{align}
    where $a = G \gamma \gamma_{\mathrm{c}} / 2$ is a scaling parameter of the measured signal corresponding to the unknown prefactor in equation~\eqref{eq:lorentzian_diff_re}.
    We fit this model to the extracted values of $\mu_{\bar S}$ with $\tilde \Delta_f = (f_{\mathrm{r},0} - f_{\mathrm p})/\alpha$, $\tilde a = a / \alpha$, and $\tilde \gamma = \gamma / \alpha$ as the fitting parameters.
    This then allows using $\mathcal{E}(s)$ as a calibration curve that maps a signal $s$ to the corresponding energy reported by the calorimeter.
    Extended Data Fig.~\ref{fig:energy_conversion}a shows $\mu_{\bar S}$ and $\mathcal{E}(s)$.
    To convert the standard deviation $\sigma_{\bar S}(E)$ at a given energy to the standard deviation $\sigma_E(E)$ which is in units of energy, we use the following relation\cite{benaroya2005} for the variance of a random variable transformed by a nonlinear function:
    \begin{align}
        \label{eq:sigma_transform}
        \sigma_E^{2}
        &
        \approx
        \mathcal{E}^2(\mu_{\bar S})
        + \sigma_{\bar S}^2\left(
            \left[\mathcal{E}'(\mu_{\bar S})\right]^2
            +
            \mathcal{E}(\mu_{\bar S})\mathcal{E}''(\mu_{\bar S})
        \right)
        -
        \left[ \mathcal{E}(\mu_{\bar S}) + \frac{\sigma_{\bar S}^2}{2} \mathcal{E}''(\mu_{\bar S}) \right]^2
        .
    \end{align}
    The standard deviations $\sigma_{\bar S}$ obtained using equation~\eqref{eq:cdf_model} and the corresponding $\sigma_{E}$ are shown in Extended Data Fig.~\ref{fig:energy_conversion}a as horizontal and vertical bars, respectively, as well as in Extended Data Figs.~\ref{fig:energy_conversion}b--c as crosses.
    The FWHM is that of a Gaussian function, given by $W_E = \sigma_E \times 2 \sqrt{2 \ln 2}$.
    The data points in Fig.~\ref{fig:calorimetry}b are thus calculated as $E_{\mathrm{MW}} / W_{E_{\mathrm{MW}}}$.

    To estimate $\sigma_E$ between the values which we have measured, we approximate $\sigma_{\bar S}(E)$ by the quadratic polynomial $\sigma_s(E) = a_0 + a_1 E + a_2 E^2$ that is fit to the measured $\sigma_{\bar S}$.
    We then transform this interpolated curve according to equation~\eqref{eq:sigma_transform} (with $s = \mathcal{E}^{-1}(E)$ in place of $\mu_{\bar S}$), resulting in the curve shown in Extended Data Fig.~\ref{fig:energy_conversion}.
    This yields the interpolated FWHM $W(E)$, and subsequently the interpolated curve in Fig.~\ref{fig:calorimetry}b, which is given by $E/W(E)$.
    Finally, the estimated energy resolution of $0.83\,\mathrm{zJ}$ is determined by numerically finding $E$ such that $E = W(E)$.

    Note that the projection angle $\tilde \varphi$, as well as the averaging window discussed above are chosen such that the energy resolution estimated with this method is optimized.
    Nevertheless, we find that for the $0.95\,\mathrm{zJ}$ pulse, even if $\tilde \varphi$ is offset by up to $0.1\times 2\pi$ from its optimal value, the measured signal-to-FWHM ratio can be made greater than unity by adjusting the averaging window offset and length by $1$--$2\,\mathrm{\mmu s}$.

\subsection*{Uncertainty of the energy resolution}

    Here, we discuss how we obtain the uncertainties of $0.02\,\mathrm{zJ}$ in the pulse energy $0.95\,\mathrm{zJ}$ and $0.04\,\mathrm{zJ}$ in the energy resolution estimate.
    It is important to make a distinction between the standard deviations $\sigma_s$ and $\sigma_E$, which stem from noise in the signal and are obtained by fitting the empirical CDFs, and their uncertainties $\delta \sigma_s$ and $\delta \sigma_E$, respectively, which are derived from the fit.
    We use \emph{uncertainty}
    to exclusively refer to $\delta \sigma_s$ and $\delta \sigma_E$, as well as to other sources of error such as the $0.1\,\mathrm{dB}$ of uncertainty in the calibration of the line attenuation.

    We obtain the uncertainty in $E/W(E)$ as follows:
    \begin{align}
        \label{eq:e_fwhm_ratio_error}
        \delta \left[\frac{E}{W(E)}\right]
        &
        =
        \frac{E}{W(E)} \sqrt{
            \left(\frac{\delta E}{E}\right)^2
            +
            \left(\frac{\delta W(E)}{W(E)}\right)^2
        },
    \end{align}
    where $\delta E$ is given by $0.1\,\mathrm{dB}\times E \approx 1.023 \times E$, and
    \begin{align}
        \delta W(E)
        =
        2\sqrt{2\ln 2}
        \,
        \delta \sigma_E(E)
        &
        =
        2\sqrt{2\ln 2}
        \,
        \sigma_s(E)
        \mathcal{E}'(s)
        \sqrt{
            \left(\frac{\delta \sigma_s(E)}{\sigma_s(E)}\right)^2
            +
            \left(\frac{\delta \mathcal{E}'(s)}{\mathcal{E}'(s)}\right)^2
        }
        ,
    \end{align}
    where
    \begin{align}
        \delta \sigma_s(E)
        &
        =
        \left[\nabla_{\mathrm{P}}\sigma_s(E)\right]^\mathrm{T}
        \mathbf{C}_{\sigma_s}
        \nabla_{\mathrm{P}}\sigma_s(E),
    \end{align}
    and
    \begin{align}
        \delta \mathcal{E}'(s)
        &
        =
        \left[\nabla_{\mathrm{P}}\mathcal{E}'(s)\right]^\mathrm{T}
        \mathbf{C}_{\mathcal{E}'}
        \nabla_{\mathrm{P}}\mathcal{E}'(s)
        ,
    \end{align}
    where $s = \mathcal{E}^{-1}(E)$ and
    $\nabla_{\mathrm{P}}\sigma_s(E)$ denotes the gradient of $\sigma_s(E)$ with respect to each of its parameters:
    $\nabla_{\mathrm{P}}\sigma_s(E) = [\partial_{a_0} \sigma_s(E),\, \partial_{a_1} \sigma_s(E),\, \partial_{a_2} \sigma_s(E)]^\mathrm{T}$,
    and similarly
    $\nabla_{\mathrm{P}}\mathcal{E}'(s)$ denotes the gradient of $\mathcal{E}'(s)$ with respect to the parameters of the Lorentzian:
    $\nabla_{\mathrm{P}}\mathcal{E}'(s) = [\partial_{\tilde a} \mathcal{E}'(s),\, \partial_{\tilde \gamma} \mathcal{E}'(s),\, \partial_{\tilde \Delta_f} \mathcal{E}'(s)]^\mathrm{T}$,
    and $\mathbf{C}_{\sigma_s}$ and $\mathbf{C}_{\mathcal{E}'}$ are the covariance matrices of the parameters, obtained from the fitting procedure described above.
    We have included the uncertainties in $\sigma_{\bar S}$ and $\mu_{\bar S}$ obtained from the CDF fitting as weights in the fitting of $\mathcal{E}$ and $\sigma_s$, so that the effect of these are included in the covariance matrices.

    Note also that the Poissonian statistics of the coherent input photons cause fluctuations in the input photon number with a standard deviation of $\sqrt{E /(h f_{\mathrm{MW}})}$, or equivalently, fluctuations in the energy with a standard deviation of $\sqrt{E h f_{\mathrm{MW}}}$.
    The effect of these fluctuations is already included in $\sigma_{\bar S}$, and hence also in the signal-to-FWHM ratio we have used to determine the energy resolution.
    For the pulse energies we have measured, this corresponds to $13$--$26$ photons, or $0.07$--$0.14\,\mathrm{zJ}$.

    The uncertainty obtained from equation~\eqref{eq:e_fwhm_ratio_error} is used to calculate the error bars and the confidence interval shown in Fig.~\ref{fig:calorimetry}b, and $\delta \sigma_s$ and $\delta \sigma_E$ are shown as the error bars and confidence intervals in Extended Data Figs.~\ref{fig:energy_conversion}b,c, respectively.
    Using these confidence intervals, we find the points where $E/W(E) \pm  \delta [E / W(E)] = 1$, which yields ($0.83 \pm 0.04)\,\mathrm{zJ}$.
    We note that the uncertainty is dominated by $\delta \sigma_s$ and $\delta \mathcal{E}'$ since the relative uncertainties resulting from the CDF fit are insignificant.

\section*{Data Availability}

The code and data that support the findings of this study are available via Zenodo at \url{https://doi.org/10.5281/zenodo.14356946}

\bibliography{zeptojoule-calorimetry}

\section*{Acknowledgements}
We thank Vasilii Vadimov and Aarne Keränen for useful discussions.
We acknowledge the provision of facilities and technical support by Aalto University at OtaNano – Micronova Nanofabrication Center and LTL infrastructure which is part of European Microkelvin Platform (EMP, no. 824109 EU Horizon 2020).
We have received funding from the European Research Council under Advanced Grant no. 101053801 (ConceptQ), Horizon Europe programme HORIZON-CL4-2022-QUANTUM-01-SGA via the project no. 101113946 (OpenSuperQPlus100), the Future Makers Program of the Jane and Aatos Erkko Foundation and the Technology Industries of Finland Centennial Foundation, Business Finland under the Quantum Technologies Industrial (QuTI) project (decision no. 41419/31/2020), the Academy of Finland through its Centers of Excellence Program (project no. 336810) and grant (project no. 349594), the Finnish Cultural Foundation, and the Vilho, Yrjö and Kalle Väisälä Foundation of the Finnish Academy of Science and Letters.

\section*{Author contributions}
The main experiments were conducted by K.K. and R.K.
The data analysis was carried out by A.M.G. and K.K., with contributions from R.K.
The calibration of the input line attenuation was carried out by J.-P.G. and K.K.
The SNS sensors used both for the main experiment and the attenuation calibration were designed and fabricated by W.L.
The manuscript was mainly written by A.M.G. with significant contributions from K.K., Q.C., J.-P.G., R.K., and M.M., and with comments from all authors.
The work was conceived and supervised by M.M.

\end{document}